\documentclass[runningheads]{llncs}
\usepackage[T1]{fontenc}
\usepackage{url}
\usepackage{soul}
\usepackage[hidelinks]{hyperref}

\title{Proving Termination and Correctness of the Partitioning Engine in the \verapak Neural Network Verifier}
\titlerunning{Termination \& Correctness of the \verapak Partition Engine}

\author{Mason Davis\inst{1}\orcidID{0009-0003-5832-8738} \and
Yongxi Sha\inst{2}\orcidID{0009-0009-4025-2660} \and
Bennett DenBleyker\inst{2}\orcidID{0000-0002-1369-6700} \and
Wen Li\inst{2}\orcidID{0000-0003-0194-2115} \and
Zhen Zhang\inst{1}\orcidID{0000-0002-8269-9489}}

\authorrunning{M. Davis et al.}

\institute{Department of Electrical and Computer Engineering, Utah State University, USA\\
\email{\{mason.davis, zhen.zhang\}@usu.edu} \and
Department of Computer Science, Utah State University, USA\\
\email{b.denbleyker@comcast.net}}

\usepackage[dvipsnames]{xcolor}
\usepackage[textsize=tiny, disable]{todonotes}
\usepackage{xargs}

\newcommandx{\md}[2][1=]{\todo[inline,
  linecolor=MidnightBlue,backgroundcolor=MidnightBlue!40,bordercolor=MidnightBlue,size=\normalsize,#1]{(Mason)
    #2}}
    
\newcommandx{\bd}[2][1=]{\todo[inline,
  linecolor=Salmon,backgroundcolor=Salmon!60,bordercolor=Salmon,size=\normalsize,#1]{(Bennett)
    #2}}
\newcommandx{\mdtobd}[2][1=]{\todo[inline,
  linecolor=CarnationPink,backgroundcolor=CarnationPink!60,bordercolor=CarnationPink,size=\normalsize,#1]{(Mason \textrightarrow\ Bennett)
    #2}}
\newcommandx{\bdtomd}[2][1=]{\todo[inline,
  linecolor=CarnationPink,backgroundcolor=CarnationPink!60,bordercolor=CarnationPink,size=\normalsize,#1]{(Bennett \textrightarrow\ Mason)
    #2}}
\newcommandx{\zz}[2][1=]{\todo[inline,
  linecolor=RawSienna,backgroundcolor=RawSienna!40,bordercolor=RawSienna,size=\normalsize,#1]{(Zhen)
    #2}}
\newcommandx{\all}[2][1=]{\todo[inline,
  linecolor=brickred,backgroundcolor=red!80,bordercolor=red,size=\normalsize,#1]{(All) #2}}

\newcommandx{\wl}[2][1=]{\todo[inline,
  linecolor=Periwinkle,backgroundcolor=Periwinkle!40,bordercolor=Periwinkle,size=\normalsize,#1]{(Wen)
    #2}}

\newcommandx{\ys}[2][1=]{\todo[inline,
  linecolor=Emerald,backgroundcolor=Emerald!40,bordercolor=Emerald,size=\normalsize,#1]{(Mark)
    #2}}
\newcommand{\verapak}{\textsc{VeRAPAk}\xspace}
\newcommand{\Verapak}{\textit{\textbf{Ve}rification for \textbf{R}obust neural networks through \textbf{A}bstraction, \textbf{P}artitioning and \textbf{A}ttac\textbf{k} methods}\xspace}
\usepackage{listings}   
\usepackage{xcolor}    
\usepackage{textcomp}

\definecolor{dkgreen}{rgb}{0,0.6,0}
\definecolor{gray}{rgb}{0.5,0.5,0.5}
\definecolor{mauve}{rgb}{0.58,0,0.82}
\definecolor{darkgray}{rgb}{0.15,0.15,0.15} 
\definecolor{purple2}{rgb}{0.87, 0.18, .92}
\definecolor{yellow}{rgb}{1, 0.8, .01}

\lstdefinelanguage{Dafny}{
  keywords=[1]{
    module, import, class, trait, datatype, newtype,
    method, function, predicate, lemma,
    if, else, while, match, case, return, yield,
    assert, assume, calc,
    requires, ensures, invariant, decreases, modifies, reads,
    is, in,
    true, false, null,
    var, const, ghost, forall, exists
  },
  keywordstyle=[1]{\color{purple2}},
  keywords=[2]{
    hierarchicalDimensionRefinement, LargestFirstDimSelection, ArgSort, PartitionRegion, Main, Array,
    IsContained, ValidRegion, RegionsDisjoint, nonEmptyRegion,
    abs, power, reverse, ArraySubtract, SequenceSubtract, productOfSeqReal, GetArea, 
    SumArea, absDiff, ItemWeight, TotalWeight, RegionBaseSeq, DimVolume, GetSplitOffset,
    LemmaSplitProperties, LemmaFractionOrder, MulNonnegative, LemmaPowerPositive, 
    LemmaMulDistributes, LemmaSubregionBounds, LemmaSumAreaAppend, 
    LemmaSequenceSubtractElement, LemmaProductBase, LemmaRegionAreaFactorization, 
    LemmaSameBaseSeq, LemmaSumAreaDistributive, LemmaSinglePartitionPreservesVolume
  },
  keywordstyle=[2]{\color{yellow}},
  keywords=[3]{
    bool, char, int, real, string, nat,
    seq, set, map, axiom, NDArray, Region, WorkItem, T
  },
  keywordstyle=[3]{\color{blue}},
  comment=[l]{//},
  morecomment=[s]{/*}{*/},
  commentstyle=\color{dkgreen}, 
  string=[b]{"},
  stringstyle=\color{mauve},
  sensitive=true
}

\usepackage{xspace}
\usepackage{amsthm, amsmath}
\usepackage{amssymb}
\usepackage{algorithm}
\usepackage{algpseudocode}
\usepackage{enumitem}

\newtheoremstyle{ieeelesson}
  {1ex}         
  {1ex}         
  {\normalfont} 
  {}            
  {\bfseries}   
  {.}           
  { .5em}       
  {\thmname{#1}\thmnumber{ #2}\thmnote{: #3}} 

\theoremstyle{ieeelesson}
\newtheorem{lesson}{Lesson}

\begin{document}

\maketitle

\begin{abstract}
Establishing trust in neural network verifiers is paramount for certifying safety-critical systems. However, these tools are inherently complex software, often relying on dynamic languages and external libraries that make them susceptible to bugs, non-terminating loops, or mathematically invalid guarantees. This paper presents a case study on formally verifying the core partitioning engine of \verapak, a modular, refutation-based neural network verifier. Because \verapak relies on swappable verification and falsification sub-routines, we isolate this central engine to establish mathematical guarantees of termination and structural correctness for the overarching framework itself. To do this, we propose a two-pronged validation methodology: First, we construct a formal model of \verapak's partitioning algorithm in Dafny, detailing practical optimization strategies for large-scale verification efforts, including isolating nonlinear arithmetic and utilizing explicit witnessing, to mitigate SMT resource exhaustion and formally prove the algorithm's correctness and guarantee its termination. Second, we introduce a contract-aware fuzzing test suite to close the semantic gap between the underlying Python implementation and the Dafny model. This provides a reusable framework for improving confidence in complex verifiers spanning multiple implementation languages.

\end{abstract}

\md{Include link to artifact in submission paper}
\clearpage
\section{Introduction}
As neural networks are increasingly deployed in safety-critical domains, the demand for neural network verification tools has grown rapidly~\cite{huang-survey-2020,Katz2017}. In response to this demand, a variety of powerful verification frameworks have been developed, including Marabou \cite{katz2019marabou,wu2024marabou}, ERAN \cite{singh2019abstract}, $\alpha,\beta$-CROWN \cite{wang2021beta}, NNV \cite{tran2020nnv,lopez2023nnv,sasaki2026n2v}, and Rocq-NN-Roll prover~\cite{rocqNNprover2026}. However, these verifiers are themselves highly complex software systems, often tasked with manipulating and evaluating high-dimensional, continuous state spaces. This inherent complexity makes them prone to underlying implementation bugs that can fundamentally compromise their reliability. If a flaw exists within a verifier's core algorithm and/or code implementation, the consequences are severe because the system might fail to terminate or could produce an invalid mathematical proof.

To establish trust in these tools, the underlying software implementation must be backed by the same mathematical guarantees. This paper presents a case study on formally verifying the core operational logic present in the \Verapak (\verapak) verification framework~\cite{verapak}. \verapak's reliance on Python and external libraries such as NumPy presents a significant barrier to formally verifying the framework itself. While Python-specific static verifiers like Nagini~\cite{eilers_nagini_2018} exist, they are not equipped to handle expansive external C-extensions like NumPy, on which \verapak heavily relies for matrix operations. Because the partitioning procedure is the core of \verapak, serving as the sole driver of search-space exploration, the difficulty in formally guaranteeing its behavior undermines the entire frameworks utility. For example, if the verification engine evaluates a region on a decision boundary, the underlying verifier may repeatedly return an ``unknown'' classification. A naive algorithm will indefinitely subdivide this boundary region in an attempt to resolve this classification, trapping the framework in a non-terminating loop. 

To bridge this gap and establish this necessary proof, we model \verapak's core $N$-dimensional ($N$-dim) partitioning procedure using the Dafny verification language~\cite{dafny}. Because \verapak is designed as a modular framework rather than a monolithic solver, it relies on swappable verification and falsification sub-routines (e.g. ERAN~\cite{singh2019abstract}) that must carry their own independent correctness guarantees. Therefore, our objective is to guarantee the termination and soundness of the framework itself.

We isolate the partitioning engine because it is the \emph{critical backbone} of the architecture and the \emph{sole driver} of the state-space exploration, dictating the termination of the overarching verification loop. By formalizing this engine, we establish the strict structural contracts required for any external verifier and falsifier to interface with \verapak. Provided these contracts are satisfied, our formalization guarantees the framework will terminate and remain sound.

Because this verified model exists independently of the framework's native execution environment, we must also address the semantic gap that exists between our idealized mathematical model operating over infinite-precision reals and the concrete Python implementation constrained by discrete floating-point arithmetic and external C-extensions. To achieve this, we introduce a \emph{contract-aware} fuzzing approach. This is not intended as a conventional fuzzing tool for crash or memory-error bug discovery. Instead, it is a complementary, property-guided testing tool. During development, it rapidly generates concrete counterexamples to highlight specification mismatches, allowing for refinement of either formal model or implementation. Once the specification is proven, the fuzzer shifts to implementation validation, increasing the confidence that these theoretical guarantees hold robustly during the concrete execution.

The major contributions of this case study are:
\begin{enumerate}
    \item \textbf{Methodologies for Large-Scale Verification}: Derived directly from the challenges of formally modeling \verapak's $N$-dim partitioning procedure, we detail practical optimization strategies for using Dafny over complex, continuous domains. This includes techniques for isolating nonlinear arithmetic, managing deeply nested loop invariants, and utilizing explicit witnessing to alleviate Z3 search burdens and improve proof stability. 

    \item \textbf{A Generalizable Fuzzing Framework}: Rather than serving as a traditional bug-finding tool, we propose a contract-aware fuzzing methodology designed for implementation validation. By enforcing preconditions on the mutator to generate sets of valid test cases, we evaluate the concrete Python execution against vectorized post-conditions oracles. This bridges the semantic gap between formal Dafny model and the native implementation, providing a reusable framework to empirically increase confidence that theoretical guarantees hold robustly at runtime. 
\end{enumerate}

\section{Background and Overview}
\label{sec:background}
\begin{figure}[h]
    \centering
    \includegraphics[width=0.77\linewidth]{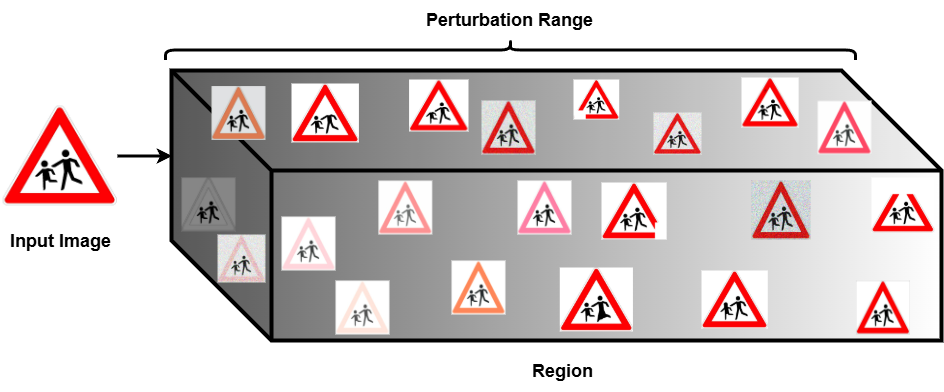}
    \caption{Illustrative visualization of an $N$-dim input region representing bounded search space around a nominal network input}
    \label{fig:input-region}
\end{figure}

\verapak is a refutation-based neural network verification framework comprising of three primary engines: verification, falsification, and partitioning. In this context, a \textit{region} is defined as an $N$-dim hyperbox representing a bounded, continuous subset of the neural network's input state space (typically encompassing allowable perturbations around a nominal input) as illustrated in Figure~\ref{fig:input-region}. A region is classified as \emph{unsafe} if it contains at least one \emph{adversarial point}, denoted as $x^*$, a valid input that violates the network's specified safety property.

Operationally, \verapak leverages its core partitioning engine to systematically prune the search space while the falsification engine rapidly identifies and eliminates unsafe regions, thereby significantly reducing the computational overhead and number of queries sent to the verification engine.

\begin{figure}[!b]
    \centering
    \includegraphics[width=\linewidth]{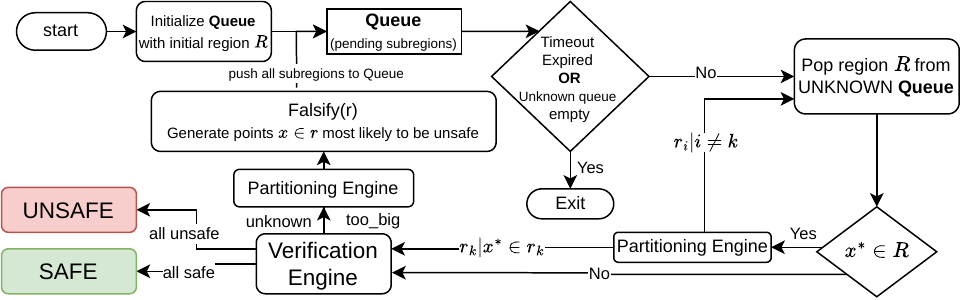}
    \caption{\verapak main verification logic flow}
    \label{fig:verapak-flow}
\end{figure}

Figure~\ref{fig:verapak-flow} illustrates the main execution loop of the \verapak framework. The algorithm operates continuously over an \texttt{UNKNOWN} queue populated initially with an $N$-dim input region. This process iterates until the queue is empty and all region area has been classified as \texttt{SAFE} or \texttt{UNSAFE}, or a timeout condition is met. During each iteration, a region $R$ is popped from the \texttt{UNKNOWN} queue. The framework first evaluates whether $R$ contains a previously discovered counterexample ($x^* \in R$). If it does, the partitioning engine subdivides $R$ to isolate the counterexample into a specific subregion, $r_k \mid x^* \in r_k$, which is then passed to the verification engine while the remaining subregions, $r_i\mid\forall i :: i \ne k$, are pushed back to the queue. 

If no counterexamples exist within $R$, the region is directly routed to the verification engine. The verifier attempts to formally classify the region as \texttt{SAFE} or \texttt{UNSAFE}. If the region is too large (memory constrained) or the verifier is unable to determine safety (returning ``unknown'' as a result), the region undergoes partitioning. These newly subdivided regions are passed through a falsification engine to generate high-probability unsafe candidate points before being returned to the queue for further processing. Through this iterative process, \verapak systematically refines the search space to isolate adversarial regions and provide high-resolution safety guarantees. 


\begin{algorithm}[t]
\caption{\verapak Hierarchical Dimension Partitioning}
\label{alg:partitioning}
\algtext*{EndFor}
\begin{algorithmic}[1]
\Require Parent region $R$ ($N$-dim), dimension selection strategy $S$, dimensions to partition $M \le N$, divisor $d > 0$
\Ensure Set of mutually disjoint, partitioned child subregions
\State $sorted\_dims \gets S(R)$ \Comment{Determine split order}
\State $V \gets \{R\}$
\For{$i = 0$ \textbf{to} $M - 1$}
    \State $dim \gets sorted\_dims[i]$
    \State $sizeIncrement \gets (R.high[dim] - R.low[dim]) / d$
    \State $V_{new} \gets \emptyset$
    \For{\textbf{each} $r \in V$}
        \For{$j = 0$ \textbf{to} $d - 1$} \Comment{Iteratively slice interval}
            \State $low \gets r.low$
            \State $low[dim] \gets low[dim] + (sizeIncrement \times j)$
            \State $high \gets r.high$
            \State $high[dim] \gets low[dim] + sizeIncrement$
            \State $V_{new} \gets V_{new} \cup \{\text{Region}(low, high)\}$
        \EndFor
    \EndFor
    \State $V \gets V_{new}$
\EndFor
\State \Return $V$
\end{algorithmic}
\end{algorithm}

We specifically target the partitioning engine for formal verification over falsification or verification engines due to its role as the critical backbone of the \verapak architecture. The verification and falsification sub-routines act as localized oracles—relying on complex, external solvers—to evaluate individual regions. In contrast, the partitioning engine is the sole driver of the framework's global state-space exploration. Its mathematical correctness is paramount: any geometric flaw, such as dropping a subregion or generating overlapping boundaries, will silently invalidate the tool's safety guarantees. Furthermore, this engine strictly dictates the termination of the overarching verification loop. If the partitioning logic fails to monotonically reduce the search space, the framework can trap itself in a non-terminating loop regardless of the underlying solvers' accuracy. Therefore, formally verifying this central engine is the most vital step in establishing end-to-end trust in the framework.

We formalize this partitioning engine in Algorithm \ref{alg:partitioning}. The core \texttt{hierarchical\-DimensionRefinement} procedure takes an initial $N$-dim parent region $R$, a dimension-selection strategy $S$ (typically ``largest-first''), the number of dimensions to partition $M$ (where $M \le N$), and a scalar divisor $d$. The algorithm operates sequentially across the selected $M$ dimensions (Lines 4-16). For each dimension, it calculates a uniform \texttt{sizeIncrement} based on the interval width and the divisor $d$ (Line 6). The engine then iterates through the current working set of regions; for every intermediate subregion, the innermost loop (Lines 9-14) performs the local subdivisions. For each of the $d$ slices, the algorithm copies the parent's boundary arrays (Lines 10, 12). It calculates the new lower bound for the active dimension by shifting it forward by an offset proportional to the current slice index $j$ (Line 11), and establishes the upper bound by adding a single \texttt{sizeIncrement} (Line 13). This newly sliced bounding box is then instantiated as a region and appended to the temporary working set $V_{new}$ (Line 14), ultimately creating exactly $d$ new child subregions per intermediate region.

\begin{figure}[!t]
    \centering
     \includegraphics[width=\linewidth]{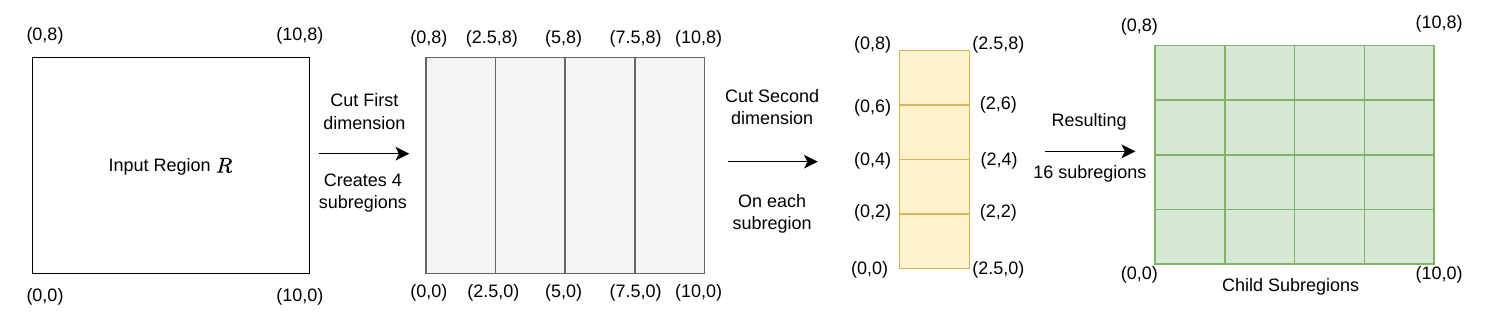}
    \caption{Toy Example: Partitioning Engine on two-dimensional region ($R$) with largest-first heuristic and divisor of 4}
    \label{fig:partitioning_example}
\end{figure}

Figure \ref{fig:partitioning_example} illustrates this operation applied to a 2D parent region $R$ with $x \in [0,10]$ and $y \in [0,8]$ using a divisor of $d=4$. Since the $x$-interval is the largest, the algorithm's outer loop first selects the horizontal dimension, and the inner subdivision loop generates four intermediate strips of width 2.5. Subsequently, the engine refines the $y$-dimension across each of those strips, applying four vertical cuts. This sequential process results in 16 uniform child subregions. The union of these resulting subregions perfectly reconstructs the original parent region $R$, ensuring that no search space is lost or duplicated during the partitioning process.


\section{Related Work}
\label{sec:relatedwork}

\ys{1. Formal verification in Dafny - suggest the limitation on counterexample generation and difficulty to interpret, so we use software fuzzing to generate meaningful counterexamples}
\noindent \textbf{Program Verification in Dafny.} While neural network verification is maturing, formal verification of verifiers remains an underexplored necessity, as implementation bugs can silently compromise their safety guarantees. Other recent advancements, such as the Rocq-NN-Roll prover~\cite{rocqNNprover2026}, verify the underlying neural network mathematics; our work solves a complementary problem by verifying the compelx software infrastructure executing those proofs. Addressing this specific infrastructural gap, recent work utilizing the Imandra automated reasoning system realized a certified proof checker for the Marabou DNN verifier~\cite{imandra-proofcheck}. However, their verification approach is execution-dependent: it relies on generating and checking proof certificates post-execution to validate the result of a single verification query. In contrast, our approach is algorithm-centric. We verify the foundational properties of the core partitioning algorithms \textit{a priori} in Dafny, establishing geometric correctness and termination for all possible execution traces before runtime, and subsequently validate the native Python implementation using contract-aware fuzzing.

Dafny provides a highly automated formal verification environment, successfully utilized in large-scale systems like Ironclad Apps ~\cite{ironclad} and IronFleet~\cite{ironfleet}. To prove correctness and termination, Dafny translates programs and their annotations into verification conditions that are automatically discharged by the Z3 SMT solver~\cite{z3} Despite this automation, writing and debugging Dafny proofs remains challenging in practice due to solver instability~\cite{mariposa} and the opacity of SMT-derived counterexamples. While prior tooling improvements have addressed trigger selection~\cite{trigger} and counterexample readability~\cite{better_counterexample}, a semantic gap persists between the verified abstract model and the concrete executable behavior. SMT-derived counterexamples are useful, but they are often partial, difficult to interpret, and frequently fail to correspond to the hardware-constrained execution of the native implementation.

\ys{2. Using Software Fuzzing to Generate Counterexamples to Support Formal Verification Implementation}

\noindent \textbf{Software Fuzzing.} Traditional mutation-based fuzzers~\cite{afl-fuzz,bohme2017directed} and property-based testing frameworks~\cite{quickcheck,klee} excel at exploring program paths, but their primary goal remains conventional bug discovery (e.g., crashes and memory errors) rather than generating structurally complex counterexamples against idealized formal specifications. Recent formal-guided fuzzers~\cite{formalfuzzer,aflnet} use models to narrow search spaces, but they do not address the continuous-to-discrete domain translation required for real-number algorithmic verification. By combining Dafny specifications with contract-aware fuzzing, our approach \emph{natively} enforces formal preconditions during input generation. This allows us to rapidly generate structurally valid test cases to validate the formal model against the concrete implementation, mitigating the limitations of standard SMT counterexamples.

\section{Modeling Challenges}
\label{sec:modelingchallengs}

Transitioning \verapak's dynamic partitioning algorithm into a mathematically rigorous environment presented several challenges and structural hurdles. Because Dafny is a statically verified language, constructing the model required careful deliberation regarding abstraction, datatype representation, and handling of complex state spaces. In this section, we explicitly state our assumptions and justify our core modeling choices.

\noindent \textbf{Challenge 1: Framework Complexity, Scope, and Interface Contracts.} 
\verapak is designed as a modular framework rather than a monolithic solver. Because our primary objective is guaranteeing the termination of the framework, we isolate the partitioning engine as the sole driver of the state-space exploration. To do this, we model the verification and falsification subroutine calls through nondeterminism. By forcing the SMT solver to evaluate the partitioning procedure across all possible classification (safe, unsafe, unknown), we guarantee the framework's soundness regardless of which external solver is currently swapped in, provided those tools carry their own independent mathematical proofs and adhere to strict interface contracts. 

We formally define the required soundness properties the external tools must satisfy to interface with \verapak's framework:

\noindent \textbf{Verification Engine:} The verification engine is a rigorous formal oracle. It is defined as a function returning a confidence value $\rho_{safe} \in [0,1]$, and optionally, a witness point $x^*$ given a region $R$ and safety property $S$. To safely interface with the framework, the engine must satisfy three soundness properties: \textbf{(1) Complete Unsafety} ($\rho_{safe}(R) = 0 \implies \forall x \in R : S(NN(x)) = \text{UNSAFE}$); \textbf{(2) Complete Safety} ($\rho_{safe}(R) = 1 \implies \forall x \in R : S(NN(x)) = \text{SAFE}$); and \textbf{(3) Witness Validity} ($\exists x^* \in \rho_{safe}(R) \implies x^* \in R \wedge S(NN(x^*))=\text{UNSAFE}$). Within our Dafny model, we abstract this continuous confidence bound into a strict classification type. The verification engine is thus modeled as a function that non-deterministically returns SAFE (mapping $\rho_{safe} = 1$), UNSAFE (mapping to $\rho_{safe} = 0$ or discovery of a valid $x^*$), or UNKNOWN (mapping to any intermediate confidence $0<\rho_{safe} < 1$).

\noindent \textbf{Falsification Engine:} The falsification engine acts as a rapid, heuristic-driven search mechanism to identify counterexamples. We define this engine as a function $F(NN, R)$ that takes a $NN$ and an input region $R$ to generate a (possibly empty) set of candidate points for safety evaluation. To interface with the framework, it must satisfy \textbf{Spatial Containment}: all generated points must lie within the region $F(NN, R) \subseteq R$. Because each candidate point is evaluated on the concrete DNN, the resulting points added to the counterexample set are provably unsafe, making all falsification engines in \verapak sound. Beyond soundness, the framework must guarantee that querying the external engines will not introduce non-terminating behavior. In \verapak, the falsification engine relies on adversarial attack algorithms (e.g., FGSM~\cite{goodfellow2014explaining}, PGD~\cite{madry2017towards}, JSMA~\cite{papernot2015limitationsdeeplearningadversarial}) and forces a strict, finite computational budget specifically limiting the number of candidate points. These single-step attacks compute a perturbation via a single forward and backward pass through the network. Because of this, their computational complexity is strictly bounded by the static architecture. Consequently, the falsification subroutine is trivially guaranteed to return control to the framework in finite time.

Within the operational \texttt{Main} loop of our Dafny model, we intentionally do not issue discrete calls to these individual subroutines. From the perspective of proving termination, both engines fulfill the exact same structural role: they either successfully classify the region (halting exploration of that branch) or return UNKNOWN (forcing further subdivision). To model this, we abstract the combined interface of both the falsification and verification engines into a unified non-deterministic branch (\texttt{if*}). This forces the SMT solver to evaluate the framework's termination measure against the absolute worst-case execution trace: the scenario where both external engines continuously return UNKNOWN, forcing the partitioning engine to exhaust its computational budget. By verifying termination under this unified abstraction, we guarantee the framework remains mathematically sound regardless of which specific engines are utilized or how frequently they fail to find a conclusive proof.

\noindent \textbf{Challenge 2: Continuous State Space Representation.} With this sound over-approximation in place, the next critical challenge lies in structurally representing the continuous state space. Because Dafny lacks a native $N$-dim array comparable to Python, our solution is to formalize a custom generic datatype that isolates dimensions from the contents: \lstinline{datatype NDArray = Array(shape: seq, data: seq)}. Crucially, we represent the \lstinline{data} as a flattened sequence rather than recursively nested sequences. Deeply nested structures in Dafny necessitate heavy use of nested quantifiers, which frequently cause excessive instantiations and divergence within the underlying Z3 SMT solver, rapidly leading to resource exhaustion and unpredictable proof timeouts~\cite{leino2016triggers}. 

This design preserves mathematical correctness without any loss of structural information, as state-space accesses in the model are exclusively required for comparing and trimming dimensions. By operating strictly on paired boundary values, i.e., seamlessly modeled as \lstinline{datatype Region = Region(low: NDArray, high: NDArray)}, the flattened arrays allow the partitioning engine to determine the width of any dimension with a single access. Because a unified linear index perfectly aligns a \lstinline{low} bound with its corresponding \lstinline{high} bound, establishing geometric correctness requires only a single quantifier over this shared index. This entirely eliminates the need for complex multi-dimensional algebraic manipulations while ensuring the verification model remains highly efficient.

\noindent \textbf{Challenge 3: Floating-Point Arithmetic Overhead.} Having addressed the structural representation of the state space, we must define the domain of the data within these regions. \verapak's native Python implementation relies on standard IEEE 754 floating-point arithmetic. While Dafny supports floating-point semantics, the underlying Z3 SMT solver evaluates them using complex bit-vector logic, which rapidly leads to severe computational overhead and unpredictable proof timeouts in high-dimensional spaces~\cite{tomb2023avoiding}. To resolve this, we instantiate the generic \lstinline{Region} boundaries using Dafny's native, infinite-precision \lstinline{real} type. This abstraction mathematically idealizes the search space, transforming discrete hardware-level complexities into a continuous over-approximation that mitigates SMT solver divergence and improves proof reliability.

Operating within this real-number domain, we formally encode \verapak's interval arithmetic by enforcing geometric invariants across the model: the \lstinline{low} and \lstinline{high} arrays must maintain equivalent shapes, and every low bound must remain strictly less than or equal to its corresponding high bound ($low[i] \le high[i]$). Together, these invariants establish the foundational correctness of the state space before partitioning. We acknowledge that utlizing reals introduces a semantic gap between our theoretical model and the discrete, floating-point Python execution. To ensure that these mathematical theorems provide robust correctness guarantees in the concrete implementation without failing due to hardware rounding, we bridge these domains with a contract-aware fuzzing methodology detailed in Section~\ref{sec:fuzzing}.

\noindent \textbf{Challenge 4: Unverified C-Extensions and NumPy.} The final and most significant barrier to verifying Python-based frameworks is their often heavy reliance on expansive C-extensions, such as the NumPy library. \verapak uses NumPy for complex array manipulation and data storage within the concrete implementation. With the $N$-dim array already implemented, we model the built-in NumPy functions on which the partitioning engine relies. The selection of dimensions along which to partition is determined by a ``largest-first" approach that makes use of the NumPy function \texttt{argsort}, returning the indices of an array in sorted order. 

Attempting to formally verify this highly optimized, low-level C codebase is computationally prohibitive, languages like Dafny are ill-equipped to track the memory manipulations, pointer arithmetic, and multi-language aspects inherent to these libraries. To address this, we define an axiomatic contract via Dafny's \texttt{\{:axiom\}} attribute that enforces what \texttt{argsort} must do, rather than how it does it. This strict specification requires that the function returns a valid, bounded permutation of indices perfectly sorted by their corresponding interval widths. By proving the downstream partitioning engine against this axiom, we guarantee the framework's correctness under the condition that the sorting utility behaves as expected, a condition we additionally validate at runtime using the proposed contract-aware fuzzing. However, while necessary for isolating the verification target, relying on axiomatic contracts introduces its own set of complexities, which are further discussed in section~\ref{sec:verif+strat}. 

\section{Termination and Correctness}
\label{sec:theorems}

With the complexities of the underlying C-extensions abstracted, we now establish the formal soundness of the core partitioning engine itself. This section presents the proof sketches and their corresponding Dafny encodings for the three critical operational guarantees: termination under finite-precision, volume preservation, and spatial disjointness.

\begin{theorem}[Termination under Finite Precision]
\label{the:termination}
    Given an $N$-dim input region $R$ with a finite interval width $W_0$, a divisor $d \ge 2$, and number of dimensions $N \ge 1$, the finite precision of IEEE floating-point arithmetic establishes a strict upper bound $K$ on the maximum possible depth of the partitioning algorithm. By mapping this theoretical bound to a discrete, strictly decreasing metric across all active subregions, the partitioning engine satisfies a well-founded decreasing measure, guaranteeing termination.
\end{theorem}

\begin{proof}
Let $W_k \le W_0 / d^k$ denote the maximum interval width after $k$ iterative subdivisions by divisor $d$. Under IEEE floating-point arithmetic, the state space is discrete with a minimum representable distance $\delta_{min} = 2^{-149}$. A boundary cannot be subdivided once $W_k \le \delta_{min}$, as bounds become adjacent floating-point numbers. Thus, the maximum subdivision depth $K$ per dimension occurs when $W_0/d^K \le \delta_{min}$. Solving this yields a finite natural number bound $K = \lceil \log_d(W_0/\delta_{min}) \rceil$. Across $N$ dimensions, the absolute maximum number of partitions is strictly bounded by $N \times K$. We map this finite physical limit to a discrete metric of remaining allowable subdivisions. Every partitioning operation strictly decrements this metric. Bounded below by zero over the natural numbers, it forms a well-founded decreasing measure, mathematically guaranteeing termination. 
\end{proof}

\textit{Encoding in Dafny:} Because Dafny's native real datatype utilizes infinite precision, a purely geometric proof of termination via volume reduction is impossible; the state space could theoretically be subdivided infinitely, resulting in non-termination. To bridge this gap, we introduce a custom datatype, \lstinline{datatype WorkItem = WorkItem(region: Region<real>, splits_left: nat)}, which pairs a state-space region with a finite \texttt{splits\_left} parameter. It is critical to note that this \texttt{splits\_left} metric is not an artificial constraint injected merely to satisfy the theorem prover, but a deliberate, mathematically derived representation of the execution environment. \verapak executes on physical hardware constrained by IEEE 754 floating-point precision. By mapping the theoretical precision limit, derived in the proof above as an absolute bound of $N \times K$, to a discrete metric, \texttt{splits\_left} bounds the infinite-precision idealized model to the finite-precision reality of the Python execution.

At the start of the framework's main execution loop (Figure \ref{fig:verapak-flow}), the initial region is instantiated with an initial budget. As established in Section \ref{sec:modelingchallengs} (Challenge 1), our Dafny specification abstracts the concrete verification and falsification engines using non-deterministic choice. By relying on this over-approximation, we force the SMT solver to verify termination across all possible execution paths, including the worst-case scenario where every subregion evaluates to \textit{unknown} and triggers maximum subdivision. 

When partitioning is required, the framework decrements the budget metric for all newly generated subregions. To formally encode the decreasing measure established in Theorem \ref{the:termination}, we define a non-compiled (or ghost) function, which calculates the remaining exploration budget across the active worklist. The SMT solver verifies that replacing a parent region with its $d^M$ children (where $M \le N$), each possessing a strictly smaller partition budget, monotonically decreases this total budget, satisfying the termination proof.

While the termination bound established in Theorem \ref{the:termination} requires explicit mathematical derivation of hardware precision limits, the geometric properties of spatial coverage and disjointness rely on standard Euclidean foundations. The primary challenge for these subsequent guarantees lies not in their mathematical derivation, but rather in encoding them scalably within an automated theorem prover. Consequently, for Theorems \ref{th:coverage+vol} and \ref{th:disjointness}, we outline the mathematical intuition and focus our detailed analysis on the structural Dafny encodings required to render these proofs tractable for the SMT solver.

\begin{theorem}[Spatial Coverage and Volume Preservation]
\label{th:coverage+vol}
    Given a valid $N$-dim input region $R$, partitioning along $M$ dimensions with a divisor $d$ yields exactly $d^M$ subregions. The geometric union of these subregions is strictly equivalent to the initial region $R$, and the sum of their individual volumes equals the original volume of $R$.
\end{theorem}


\textit{Proof Sketch:} We approach Theorem \ref{th:coverage+vol} by induction on the sequence of partitioned dimensions. Because a single-axis split strictly divides the target interval into exactly $d$ adjacent segments without gaps or overlaps, leaving all unpartitioned dimensions unchanged, the combined geometric volume perfectly equals the parent. Inductively applying this across $M$ dimensions yields exactly $d^M$ total subregions while preserving total volume.

\textit{Encoding in Dafny:} We translate this mathematical guarantee into a strict method contract for the core partitioning engine. By defining the volume geometrically via the \texttt{GetArea} and \texttt{SumArea} functions, Dafny verifies that no search space is lost during this refinement procedure. Figure \ref{fig:volumepresinner} highlights the foundational postconditions for the inner-most loop (Algorithm \ref{alg:partitioning}, Lines 9–14), which isolates partitioning to a single dimension. Here, we prove the sum of the children's 1D volume (\texttt{DimVolume}) along the cut dimension $dim$ exactly equals the parent's width. Additionally, we verify that the boundaries of all untouched spatial dimensions (represented by the index $k$ in the Dafny specification) remain identical to the parent, ensuring the cross-sectional area is preserved for the overall geometric volume calculation.

\begin{figure*}[!b]
    \centering
    \begin{lstlisting}[language=Dafny]
method PartitionRegion(parent: Region<real>, dim: int, divisor: int) returns (children: seq<Region<real>>)
ensures forall c :: c in children ==> 
|c.low.data| == |parent.low.data| == |c.high.data| &&
(forall k :: 0 <= k < |parent.low.data| && k != dim ==> 
    c.low.data[k] == parent.low.data[k] && c.high.data[k] == parent.high.data[k])
ensures DimVolume(children, dim) == parent.high.data[dim] - parent.low.data[dim]
    \end{lstlisting}
    \caption{Volume Preservation Contract: innermost subdivision loop (Algorithm \ref{alg:partitioning}, 9--14)}
    \label{fig:volumepresinner}
\end{figure*}

\begin{theorem}[Spatial Disjointness]
\label{th:disjointness} Let $R$ be an $N$-dim parent region. For any two distinct child subregions $A$ and $B$ generated from $R$ during the partitioning procedure, their geometric intersection contains zero internal volume. They are strictly mutually disjoint, sharing at most their hyper-rectangular dimensional boundaries.
\end{theorem}


\textit{Proof Sketch:} We establish Theorem \ref{th:disjointness} by proving that the partitioning engine actively maintains a strictly non-overlapping interval along at least one dimension for any two distinct subregions. Specifically, the inner loop (Algorithm \ref{alg:partitioning} Lines 9-14) divides a target dimension into $d$ adjacent segments via monotonic offsets, mathematically constraining the $i$-th child's upper bound to never exceed the $(i+1)$-th child's lower bound.

This 1-dimensional separation is the basis of the proof. Because an $N$-dim region is defined geometrically as the Cartesian product of its 1-dimensional intervals, two regions can only share internal volume if they strictly overlap across \emph{every} dimension. By preserving our 1-dim sequential slicing across the $N$-dim state space, we guarantee that for any two final subregions $A$ and $B$, there exists at least one dimension $k$ where they are geometrically separated ($A_{high}[k] < B_{low}[k]$ or $B_{high}[k] < A_{low}[k]$). Because their intersection evaluates to a width of zero along the $k$ axis, the product of their combined $N$-dim volume is forced to zero. Consequently, regardless of how heavily these subregions overlap in the remaining $N-1$ dimensions, it is impossible for them to share internal volume, ensuring they are strictly disjoint.

\textit{Encoding in Dafny:} We formalize this property by defining the \texttt{Regions\allowbreak Disjoint} predicate to evaluate two $N$-dim regions. Utilizing an existential quantifier (\texttt{exists k}), the predicate asserts there is at least one dimension index where the boundaries strictly separate the regions ($r1.high[k] \le r2.low[k]$ or $r2.high[k] \le r1.low[k]$). Within the innermost subdivision loop of the \texttt{Partition\allowbreak Region} method, Dafny verifies that the derived monotonic offsets produce adjacent, non-overlapping bounds. This is continuously enforced as a strict loop invariant (\lstinline{forall i, j :: 0 <= i < j < d ==>}\allowbreak \lstinline{RegionsDisjoint(children[i], children[j])}) where d is current offset, explicitly proving that the subregions generated along this single axis share no internal volume.



After establishing the properties of cardinality, spatial coverage, volume preservation, and disjointness, for the innermost loop (Algorithm \ref{alg:partitioning}, Lines 9--14), we compose these guarantees across the iterative $N$-dim refinement process, culminating in a comprehensive method contract for the primary execution of the engine, \texttt{hierarchicalDimensionRefinement}, illustrated in Figure \ref{fig:final-contract}. While these theorems validate the geometric soundness of individual 1D cross-sections, this overarching specification guarantees that the composition of these linear partitions remains sound across the entire domain. Ultimately, by enforcing these properties as postconditions on the primary execution method, the SMT solver verifies that the framework \textit{neither abandons viable search space nor introduces redundant, overlapping subregions}.

\begin{figure}[t]
    \centering
    \begin{lstlisting}[language=Dafny, framesep=2pt, aboveskip=0pt, belowskip=0pt]
method hierarchicalDimensionRefinement(region: Region<real>, num_dimensions: nat, divisor: int) 
    returns (retVal: seq<Region<real>>)
  // Pre-conditions: Valid input geometry and valid parameters
  requires nonEmptyRegion(region) && ValidRegion(region) && divisor > 0
  requires num_dimensions <= |region.low.data| && num_dimensions <= |region.high.data|
  // 0. Validity: All generated subregions remain geometrically valid
  ensures (forall r :: r in retVal ==> ValidRegion(r))
  // 1. Cardinality: The exact number of generated subregions is d^M
  ensures |retVal| == power(divisor, num_dimensions)
  // 2. Spatial Containment: Every subregion is bounded by the parent
  ensures (forall r :: r in retVal ==> |r.low.data| == |region.low.data|)
  ensures (forall r :: r in retVal ==> |r.high.data| == |region.high.data|)
  ensures (forall r :: r in retVal ==> IsContained(r, region))
  // 3. Mutual Disjointness: No two subregions overlap in internal volume
  ensures (forall r1, r2 ::r1 in retVal && r2 in retVal && r1!=r2 ==> RegionsDisjoint(r1, r2))
  // 4. Volume Preservation: Sum of subregion volumes exactly equals parent volume
  ensures SumArea(retVal) == GetArea(region)
    \end{lstlisting}
    \caption{Contract for the $N$-dim partitioning engine}
    \label{fig:final-contract}
\end{figure}

\section{Verification Strategies and Lessons Learned}
\label{sec:verif+strat} 

Translating the high-level partitioning procedure of \verapak into a mathematically rigorous Dafny model presented significant challenges that extended beyond initial datatype abstractions. Formally reasoning about $N$-dim input spaces exposed an inherent brittleness of automated SMT solving. Specifically, the combination of infinite-precision reals, deeply nested loops and their invariants, and quantification triggers for our axiom abstractions frequently resulted in solver timeout or inconsistent verification results. To make the verification tractable and to ensure proof stability, we developed a series of strategies that are outlined in this section, offering a reproducible blueprint for verifying complex algorithms in Dafny. 

\begin{lesson}[Nested Loops and Invariants]
Verifying $N$-dim iterative operations naturally leads to deeply nested loops. In our Dafny model, these are specifically represented as \texttt{while} loops. Passing state context from an outer loop into an inner loop requires exhaustive, compounding invariants. Rather than unrolling complex loop body logic, Dafny forces the SMT solver to reason across iterations using only these explicit loop invariants. Because of this, verifying the full $N$-dim partitioning pipeline with its three nested loops leads to solver timeouts. Furthermore, these deeply nested invariants quickly become burdensome, increasing the risk of introducing conflicting or competing solver constraints. 

To help mitigate this, we modularized the pipeline. By extracting the innermost loop, which handles the complex splitting logic across one dimension, into an isolated method, we reduced the complexity of invariants needed. This required that the outer loop invariants implied the precondition, establishing the necessary information for the innermost loop's work, while the innermost loop invariants now only had to imply the postcondition which implied the outer loop invariant. This modularization removes the invariant dependency at the deepest point which saw improved performance for the underlying SMT solver. 

From this process, we identified a generalizable, bidirectional methodology for verifying nested loops in Dafny. Mathematically, the innermost loop invariants must ultimately be the strongest, logically implying the invariants of the outer enclosing loops. Since the innermost loop handles direct state modifications, defining its core correctness invariants is generally straightforward. Therefore, it is highly effective to begin at this inner level and work outward, constructing invariants that track the cumulative work done inside. Additionally, it is vital to adopt a top-down perspective to identify necessary state context, explicitly specifying what information the outer loops must carry forward into the inner scope. Through an iterative process of adjusting and tightening these constraints, a minimal and optimal invariant list can be achieved. 
\end{lesson}

\begin{lesson}[Nonlinear Arithmetic]
The core partitioning heuristic requires dividing a continuous interval width by a scalar divisor. In Dafny, multiplication or division across continuous variables introduces nonlinear arithmetic. The underlying Z3 SMT solver natively struggles with nonlinear real arithmetic, frequently resulting in unstable heuristics and unpredictable search spaces. This presents itself as brittle proofs within Dafny, seemingly verifying in one instance, only to fail after adding or changing a non-related area of the code~\cite{tomb2023avoiding}. This is further compounded within deeply nested loop invariants, causing timeouts and high computational overhead, rendering these proofs intractable and highly brittle. 

To resolve this, we employ two structural mitigation strategies. First, we isolate and encapsulate all nonlinear arithmetic using Dafny's \texttt{\{:opaque\}} attribute. In our proof, this results in a function to get the split-offset for the partitioning of a single dimension which is shown in Figure \ref{fig:nonlinear-strategies}. Here, we multiply the interval width, \texttt{total\_diff}, by the current slice to determine the new boundaries. By declaring \texttt{GetSplitOffset} as opaque, we force Dafny to expose only the function's signature to the global solver context, completely hiding the underlying nonlinear division and multiplication. This reduces the overall expensive nonlinear queries from the search space. When the arithmetic is required to prove geometric boundaries, we expose it using an explicit \texttt{reveal} statement within a dedicated, and strictly scoped lemma illustrated in Figure~\ref{fig:nonlinear-strategies}.


Second, we establish a foundation for basic continuous arithmetic. While Z3 efficiently deduces simple mathematical properties over integers, it frequently fails to autonomously generalize these axioms to the real domain especially, when presented as nonlinear arithmetic. To ease the solver's search burden and prevent timeouts, we explicitly construct helper lemmas to verify foundational algebraic properties. For instance, to successfully reason about the disjointness of our partitioned subregions, we explicitly define the preservation of inequalities under scalar multiplication shown in Figure \ref{fig:nonlinear-strategies}.


\begin{figure}[!t]
    \begin{lstlisting}[language=Dafny, basicstyle=\footnotesize\ttfamily]
// 1. Nonlinear Arithmetic Encapsulation
function {:opaque}
GetSplitOffset(total_diff: real, k: int, divisor: int): real
  requires divisor > 0 && total_diff >= 0.0
{ total_diff * (k as real / divisor as real) }

// 2. Split properties using reveal
lemma LemmaSplitProperties(total_diff: real, k: int, divisor: int)
  requires divisor > 0 && total_diff >= 0.0 && 0 <= k < divisor
  ensures GetSplitOffset(total_diff, k, divisor) <= 
          GetSplitOffset(total_diff, k+1, divisor)
{ reveal GetSplitOffset(); }

// 3. Foundational algebraic properties
lemma MulNonnegative(a: real, b: real, c: real)
  requires a >= 0.0 && b <= c
  ensures a * b <= a * c
{}
    \end{lstlisting}
    \caption{Strategies for Nonlinear Arithmetic: Opaque encapsulation, scoped reveal, and foundational algebraic lemmas.}
    \label{fig:nonlinear-strategies}
\end{figure}

By providing these primitive mathematical building blocks directly, we can prevent the solver from stalling or failing on fundamental arithmetic, making the overarching verification both tractable and stable. Verifying continuous domains in automated theorem provers requires shifting the computational burden away from the solver's internal heuristics and onto the structural design of the proof itself. The key takeaways to working with nonlinear arithmetic and the real space is twofold: First, isolate the nonlinear arithmetic using opaque functions to protect the global search space. Second, never assume the SMT solver will infer basic algebraic properties. By explicitly defining these foundational axioms as helper lemmas, verification engineers can maintain control over the proof's stability, predictability, and lower execution time.

\end{lesson}

\begin{lesson}[Quantifiers and Triggers]
To make the verification of \verapak's partitioning engine tractable, we abstracted its reliance on NumPy functions, specifically \texttt{argsort}, using Dafny's \texttt{\{:axiom\}} attribute. This allowed us to define the strict sorting contract without verifying the underlying C implementation. However, employing axiomatic contracts for complex array operations introduces a subtle risk. Because an axiom provides no executable code body, the SMT solver relies entirely on the postconditions to understand its behavior. Formally specifying the results of a sorting algorithm inherently requires extensive use of universal (\texttt{forall}) quantifiers. The underlying SMT solver evaluates these quantified statements using a pattern-matching heuristic.

To instantiate a quantified statement, the solver requires a specific syntactic pattern, or \textit{trigger} in Dafny, to appear in the localized proof context. If the expression is too complex or ambiguous, Dafny may fail to infer a valid trigger. Consequently, the SMT solver will silently omit the quantified clause from the query. This will cause any proof relying on the axiom's property to fail unpredictably, often resulting in spurious counterexamples. When Dafny's automatic trigger inference fails, often within complex sequence comprehensions or external axioms, proof stability requires manual intervention. Engineers can provide unambiguous syntactic patterns via explicit \texttt{\{:trigger\}} annotations, or avoid over-engineering the axiom by forcing instantiation directly at the call site. We adopted the latter approach, restating critical postconditions as explicit \texttt{assert} statements immediately following the function call. Both strategies (detailed in \ref{app:triggers}) guarantee axiom instantiation, eliminating a major source of verification brittleness.
\end{lesson}

\begin{lesson}[Explicit Witnessing]
In automated theorem proving, existential quantification is computationally expensive. When evaluating dynamically generated $N$-dim subregions, establishing the parent-child relationship is critical for proving structural properties like geometric containment. A naive specification attempts to enforce this by asserting that for every generated subregion, there exists some parent in the previous worklist that geometrically contains it. However, applying this existential quantification over high dimensional, continuous space forces the SMT solver into a massive combinatorial search. This floods the underlying search space and leads to extreme verification times or timeout. 

To eliminate this search burden, we replaced existential quantification with ``explicit witnessing'' through the use of Dafny's \texttt{ghost} variables. In Dafny, \textit{ghost state} refers to variables and constructs that are strictly utilized for proof and specification. Because they are completely erased by the compiler, they incur absolutely no computational overhead in the final compiled executable; for a comprehensive introduction to these concepts and further examples, we direct the reader to~\cite{Leino2023-progproofs}. Rather than forcing the solver to discover the parent-child relationship post-generation, we manually track this lineage during the execution of the loop. By introducing a ghost variable, we explicitly record the exact index of the parent alongside the generation of the child. Figure \ref{fig:explicit-witnessing} illustrates this transformation, contrasting the naive existential search with our optimized explicit lookup. This transforms the existential search into a simple array access. The solver no longer needs to evaluate \textit{if} a valid parent exists; it is handed the witness and only needs to verify the containment property for that pair.

\begin{figure}[!b]
    \centering
    \begin{lstlisting}[language=Dafny]
// --- Existential Quantification ---
invariant forall c :: 0 <= c < |newRetVal| ==> 
    exists p :: 0 <= p < |retVal| && IsContained(newRetVal[c], retVal[p])
// --- The Explicit Witness Approach ---
ghost var parentMap: seq<int> := [];
...
invariant forall k :: 0 <= k < |newRetVal| ==> 
    0 <= parentMap[k] < |retVal| && 
    IsContained(newRetVal[k], retVal[parentMap[k]])
    \end{lstlisting}
    \caption{Explicit witnessing via ghost state}
    \label{fig:explicit-witnessing}
\end{figure}

Improving verification time and proof stability is vital. Whenever a framework iteratively generates or transforms collections of data, relying on the solver to retroactively rediscover relational properties, such as lineage, mapping, or partitioning, inevitably invites combinatorial explosion. The broader implication of our explicit witnessing strategy is that it provides a highly generalizable solution to this problem: \emph{shifting the discovery burden away from the solver and into the structural specification}. Because ghost state incurs no computational overhead, verification engineers can freely embed rich relational data directly into operational loops. By proactively supplying these manual witnesses, what would otherwise be an exponential, intractable search space is systematically transformed into a stable, linear verification task.
\end{lesson}

\section{Closing the Semantic Gap: Contract-Aware Fuzzing}
\label{sec:fuzzing}

While our Dafny model provides absolute mathematical guarantees over the abstract logic, the concrete execution of \verapak relies on Python, external extensions, and discrete floating-point arithmetic. To bridge this semantic gap, we developed a contract-aware fuzzing test suite using Atheris~\cite{atheris_github}, a coverage-guided fuzzing engine. Rather than treating the fuzzer as a black box, we leverage the Dafny specifications to dictate input generation and output validation. 

\noindent \textbf{Precondition-Guided Generation.} 
To ensure input mutation provides increased coverage while strictly enforcing the Dafny model's preconditions, we create a custom, precondition-guided mutator. It operates directly on the deserialized \texttt{Region} objects, applying a randomly selected geometric transformation such as boundary sliding, expansion, or shrinking to any number of dimensions. Crucially, the mutator is \emph{contract-aware}, forcing the adherence to the preconditions prior to execution. By embedding these formal conditions into the generation phase, we guarantee the fuzzer only produces valid inputs for downstream validation.

\noindent \textbf{Postcondition Oracles.} Once a valid input is generated, it is executed against the native Python codebase function of interest. To evaluate correctness of the resulting concrete outputs, we construct test oracles derived directly from the Dafny postconditions. However, translating continuous, infinite-precision real constraints into bit-precise Python assertions yields false positives due to IEEE 754 floating-point rounding errors. Briding this mathematical domain mismatch, our oracles evaluate constraints using hardware-appropriate tolerances (e.g., relative $R_{tol}=1e^{-4}$ and absolute $A_{tol}=1e^{-5}$ via \texttt{np.isclose}). This strictly enforces the formal intent while accommodating permissible precision drift.

Translating Dafny's formal postconditions into executable Python oracles requires mapping symbolic quantifiers to highly optimized array operations. To verify properties like spatial containment and mutual disjointness across $N$ dimensions, the oracle avoids slow, deeply nested loops. Instead, these first-order logic statements are translated into vectorized NumPy operations, utilizing combinatorial iterators and element-wise boolean arrays aggregated via \texttt{np.all}. This direct translation (detailed in \ref{app:oracles}) ensures the oracle remains mathematically faithful and computationally efficient during high-throughput fuzzing.


\noindent \textbf{Validation of Axiomatic Justification.} To ensure our abstraction of NumPy's \texttt{argsort} is mathematically sound, the fuzzer evaluates the concrete dimension selection logic against an oracle mirroring the formal \texttt{\{:axiom\}} contract. This checks for: (1) correct dimensional cardinality, (2) valid index bounds, (3) dimension uniqueness, and (4) strict ``largest-first" sorting of interval widths. Any violation is captured as a concrete counterexample to refine either the model or the implementation.

Finally, we apply this fuzzing technique to the core partitioning engine itself to empirically validate the primary theorems established in Section \ref{sec:theorems}. As the precondition-guided mutator continuously generates highly irregular, yet structurally valid regions, a Python-native oracle evaluates the resulting subregions produced by the engine. The oracle asserts that the concrete execution maintains the five geometric properties (as seen in Figure \ref{fig:final-contract}): (0) geometric validity, (1) cardinality, (2) spatial containment, (3) mutual disjointness, and (4) volume preservation. By systematically validating the concrete execution against these proven specifications, we establish a high degree of confidence that \verapak's theoretical guarantees hold robustly at runtime.

\section{Evaluation and Results}
\label{sec:results}
To validate the utility of our dual-pronged methodology, we evaluate the performance of the Dafny model and the efficacy of our contract-aware fuzzer. To facilitate reproducibility, the complete Dafny formalization of the partitioning engine and the contract-aware fuzzing suite are included in the artifact archived on Zenodo~\cite{artifact2026}. All experiments were executed on an Intel Core i7-14700K (64 GB RAM, Windows 11). 

Evaluation of the formal model compares execution logic against annotation overhead. Here, our Dafny model's core logic, including datatypes, dimension selection axioms, and the primary \texttt{hierarchical\-DimensionRefinement} loops, is compact, requiring approximately 100 lines of code. However, mathematically isolating the nonlinear arithmetic and proving geometric properties over $N$-dim space requires substantial scaffolding. Auxiliary lemmas, explicit ghost-state witnessing, and strict loop invariants bring the total model size to roughly 400 lines of code. The complete Dafny source code and corresponding formal specifications are publicly available on GitHub~\cite{dafnycode-repo2026}.

Despite this annotation overhead, our structural mitigation strategies successfully rendered the continuous-domain proof tractable. By encapsulating the computationally expensive operations and explicitly guiding the solver's existential quantification, the solver verifies the entire formalized framework in 2 minutes and 8 seconds. This demonstrates that the techniques outlined in this case study can serve as a reproducible blueprint for scaling automated theorem proving to high-dimensional, continuous state spaces. By proactively managing the SMT solver's search burden, engineers can establish guarantees over highly complex software frameworks without facing unpredictable timeouts and computational costs. 

\begin{figure}[!t]
    \centering
    \begin{minipage}[t]{0.48\textwidth}
        \centering
        \includegraphics[width=\linewidth]{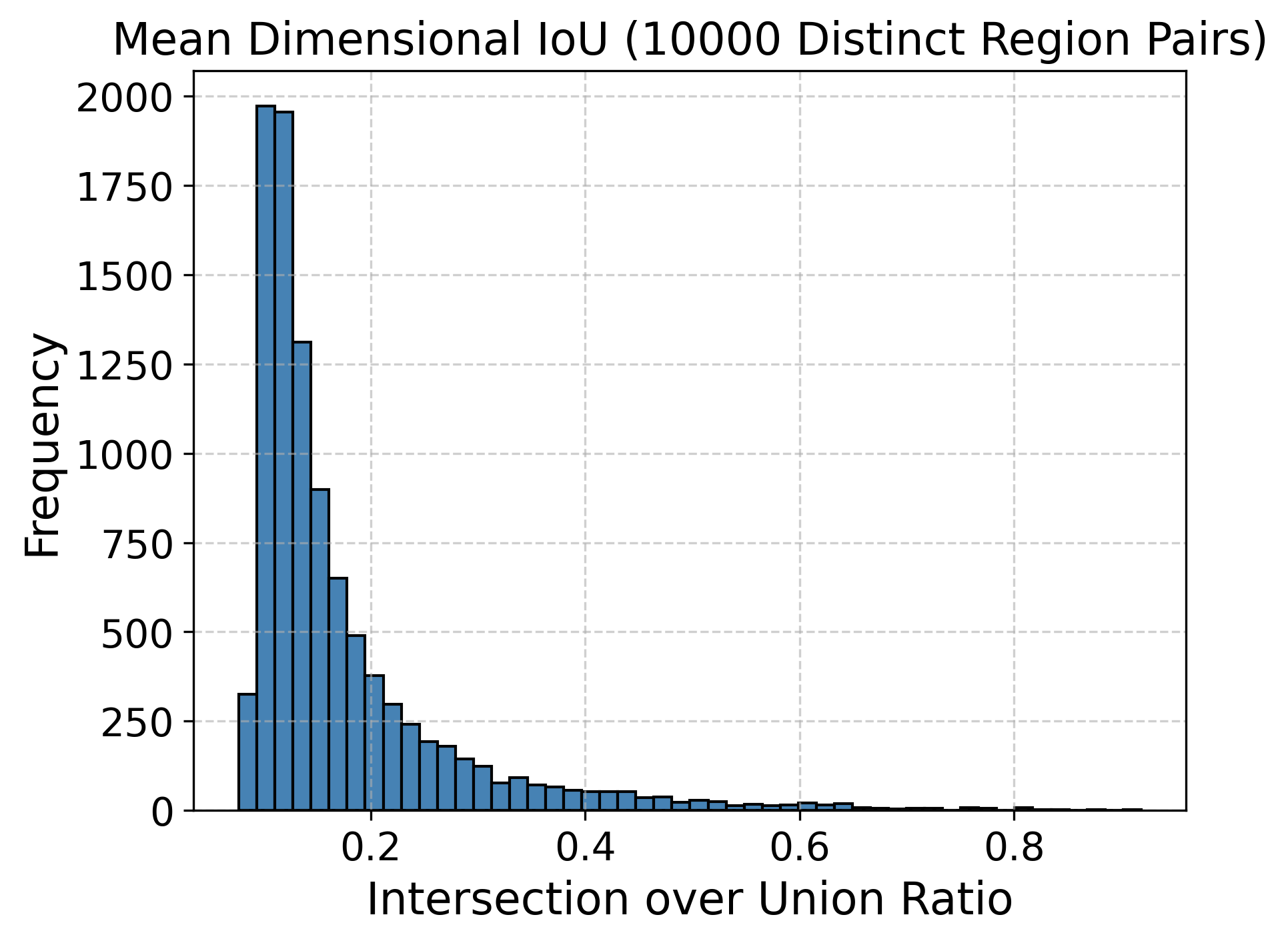}
        \vspace{-1em} 
        \caption{Frequency distribution of mean dimensional IoU across 10,000 distinct generated region pairs.}
        \label{fig:iou}
    \end{minipage}\hfill
    \begin{minipage}[t]{0.48\textwidth}
        \centering
        \includegraphics[width=\linewidth]{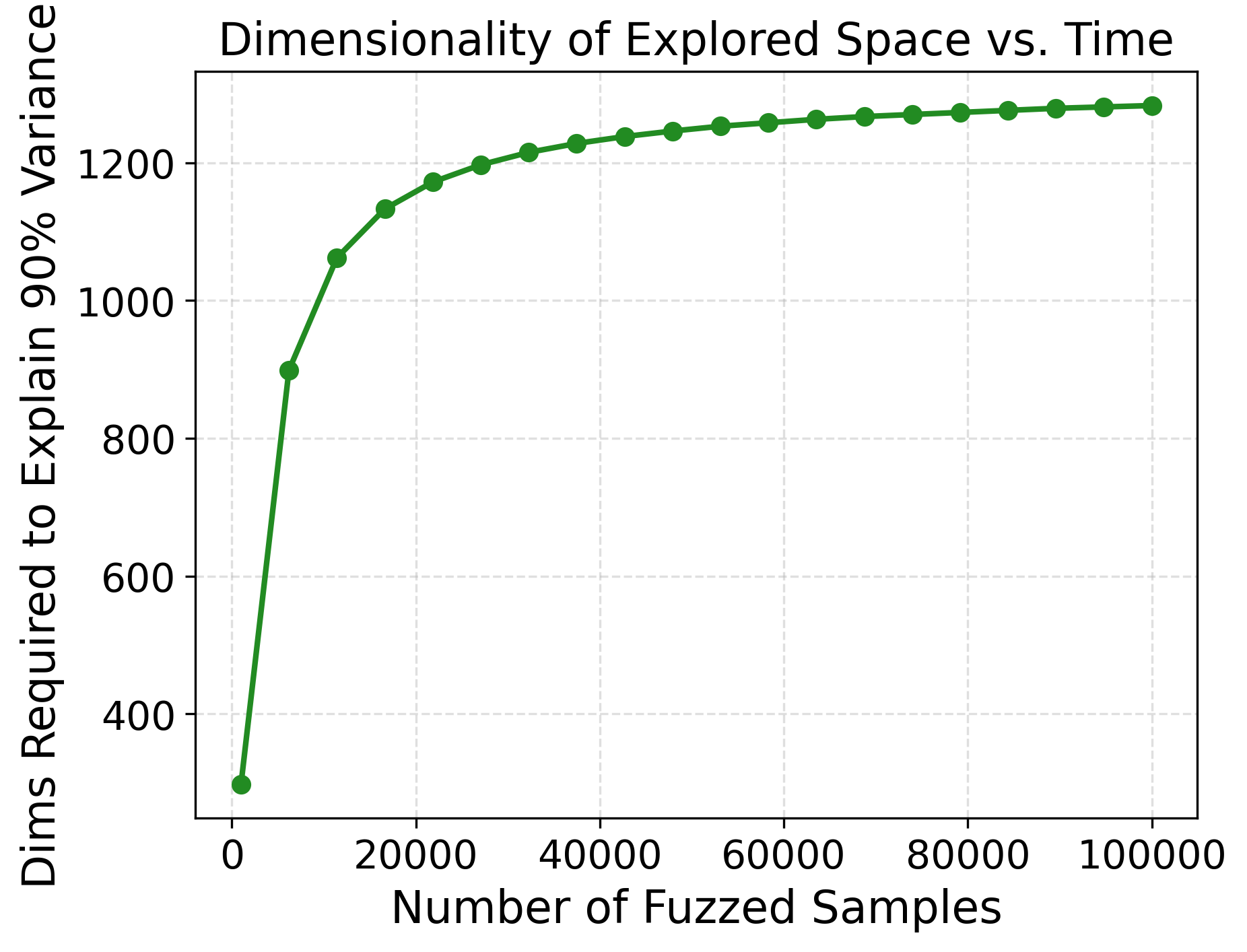}
        \vspace{-1em} 
        \caption{Dimensionality required to explain 90\% variance mapped across the fuzzing timeline.}
        \label{fig:pca-time}
    \end{minipage}
\end{figure}

Evaluating the Python codebase, our precondition-guided mutator generated 100,000 $N$-dim permutations. Because standard code-coverage percentages fail to capture input diversity in localized state-space exploration, we mathematically quantify our search using two specialized metrics: Intersection Over Union (IoU) AND Principal Component Analysis (PCA). To measure spatial overlap, we compute the mean 1-dim IoU, evaluating the intersection width divided by the union width element-wise, across 10,000 distinct, randomly sampled region pairs. As illustrated in Figure~\ref{fig:iou}, the pairs share an average of less than 20\% overlap. This low average indicates a mathematically diverse, non-repetitive search, confirming the mutator successfully avoids localized geometric traps.

To provide a mathematical measure of structural diversity, we applied PCA directly to the flattened region boundaries. PCA transforms a dataset into orthogonal principal components ordered by variance; a poorly mutated, repetitive input space would require very few components to explain its variance. Evaluating a standard $28\times28\times1$ MNIST benchmark (which flattens into 1568 continuous boundary dimensions), our analysis reveals that explaining 50\%, 90\%, and 99\% of variance requires 532, 1283, and 1537 dimensions, respectively. This confirms the generated input space is changing drastically and exploring a massive structural variety.

Finally, we utilize this PCA to mathematically justify our execution timeline. Figure~\ref{fig:pca-time} tracks the number of required dimensions to explain 90\% of the variance across 20 distinct milestones during the fuzzing process. The dimensionality reaches a clear asymptote as the sample count approaches 100,000. Running the fuzzer beyond this point yields diminishing returns, as structual diversity does not increase significantly. 

Given this diverse state-space generated by the mutator, achieving \emph{zero} post-condition oracle violations across all 100,000 iterations for each function provides a high degree of empirical confidence that the discrete Python implementation faithfully adheres to the formal Dafny specifications. While traditional software fuzzing is evaluated by the quantity of crashes, vulnerabilities discovered, or code coverage, finding zero violations in this context is successful. This yields two critical insights. First, it provides increased confidence \verapak's core implementation mirrors the specifications proven in the formal Dafny model. Second, it shows that our domain-bridging strategy is effective. This allows for translation between Dafny's real domain to Python's hardware limited floating-point domain without triggering false logic violations from floating-point rounding. 

Ultimately, coupling these continuous-space proofs with contract-aware fuzzing provides a generalizable blueprint for establishing rigorous guarantees in other complex, multi-language software frameworks. This dual-pronged approach demonstrates how engineers can scale automated theorem proving to high-dimensional, continuous state spaces while maintaining high empirical confidence in the underlying, hardware-constrained discrete implementations.

\section{Conclusion}
\label{sec:conclusion}

The reliability of neural network verification is fundamentally limited by the correctness of the verifiers themselves. By formally modeling \verapak's $N$-dim partitioning engine in Dafny and implementing a contract-aware fuzzing suite, we have established a scalable blueprint for bridging the gap between idealized mathematical proofs and concrete Python execution. This two-pronged methodology not only proves the geometric correctness and termination of the core framework but also provides reproducible strategies for stabilizing SMT-based proofs in continuous domains. Future work will focus on extending this contract-constrained validation by automating the generation of contract-aware mutators directly from Dafny specifications. This will further streamline the formal validation of high-dimensional, safety-critical software.

\bibliographystyle{splncs04}
\bibliography{references.bib}

\clearpage
\appendix
\renewcommand{\thesection}{Appendix \arabic{section}}
\section{Managing Axiomatic Triggers in Dafny}
\label{app:triggers}

As discussed in Section \ref{sec:verif+strat}, abstracting complex C-extensions like NumPy's \texttt{argsort} requires axiomatic contracts. However, because axioms lack an executable body, the SMT solver relies entirely on the specified postconditions. A significant challenge arises when these postconditions rely on universal quantifiers (\texttt{forall}). The solver evaluates these using a pattern-matching heuristic. If the solver cannot infer a valid trigger from the syntax, it may silently ignore the clause, leading to spurious counterexamples. 

Verification engineers can resolve this trigger inference failure using two primary strategies:

\textbf{Strategy 1: Explicit Wrappers and Annotations.} Engineers can define a simple wrapper function (e.g., an \texttt{Identity} function) and use it to manually annotate the quantifier with a \texttt{\{:trigger\}} statement  (Figure \ref{fig:trigger-wrapper}). This provides the solver with an unambiguous syntactic pattern to match against.

\begin{figure}[h]
    \centering
    \begin{lstlisting}[language=Dafny]
function Identity(x: int): int { x }

method {:axiom} ArgSort_Wrapper(s: seq<real>) returns (result: seq<int>)
  // The solver now has trigger pattern: Identity(i)
  ensures (forall i :: {:trigger Identity(i)} 
          0 <= Identity(i) < |s| ==> 0 <= result[Identity(i)] < |s|)
    \end{lstlisting}
    \caption{Strategy 1: Using an explicit opaque wrapper function to provide a clear trigger pattern.}
    \label{fig:trigger-wrapper}
\end{figure}

\textbf{Strategy 2: Call-Site Instantiation.} 
To avoid modifying the axiom's internal structure, engineers can manually force instantiation at the call site (Figure \ref{fig:trigger-callsite}). By restating the critical axiomatic postconditions as explicit \texttt{assert} statements immediately following the function call, the solver is forced to instantiate the universal quantifier for those specific local variables.

\begin{figure}[h]
    \centering
    \begin{lstlisting}[language=Dafny]
method LargestFirstDimSelection(region: Region<real>) 
    returns (largest: seq<int>)
{
    var diff := absDiff(region);
    var smallest := ArgSort(diff); 
    
    // assert pulls the sortedness guarantee into 
    // local proof context
    assert(forall i, j :: 0 <= i < j < |diff| ==> 
            diff[smallest[i]] <= diff[smallest[j]]);
}
    \end{lstlisting}
    \caption{Strategy 2: Forcing quantifier instantiation via explicit call-site assertions.}
    \label{fig:trigger-callsite}
\end{figure}

\section{Translating Formal Contracts to Vectorized Oracles}
\label{app:oracles}

Section \ref{sec:fuzzing} introduces our contract-aware fuzzing methodology, which requires bridging the semantic gap between Dafny's infinite-precision, first-order logic and Python's hardware-constrained, discrete execution. To maintain the computational efficiency required for high-throughput fuzzing, the Dafny sequence quantifiers must be translated into vectorized NumPy operations rather than nested Python \texttt{for}-loops. 

Furthermore, direct bit-precise comparisons (e.g., \texttt{child.low >= parent.low}) yield false positives due to floating-point arithmetic drift. To compensate, the Python oracles incorporate relative ($R_{tol}=1e^{-4}$) and absolute ($A_{tol}=1e^{-7}$) tolerances using \texttt{np.isclose}.

Figure \ref{fig:oracle-translation} demonstrates this direct translation for the Spatial Containment property. 

\begin{figure}[h]
    \centering
\begin{lstlisting}[language=Python]
# Spatial Containment Translation
# Dafny Contract: 
# forall k :: r_outer.low.data[k] <= r_inner.low.data[k] <= 
#             r_inner.high.data[k] <= r_outer.high.data[k]
# =======================================
low_check = (child.low > parent.low) | np.isclose(child.low, parent.low, rtol=RTOL, atol=ATOL)
contained_low = np.all(low_check)

high_check = (child.high < parent.high) | np.isclose(child.high, parent.high, rtol=RTOL, atol=ATOL)
contained_high = np.all(high_check)

is_contained = contained_low and contained_high
\end{lstlisting}
    \caption{Translation of spatial containment contract into a vectorized Python oracle.}
    \label{fig:oracle-translation}
\end{figure}

\end{document}